\documentclass[12pt]{amsart}
\usepackage{epsfig}
\usepackage{amsmath}
\usepackage{amssymb}

\def\grafa#1{\begin{center} \epsfig{file=#1, width=7.7cm} \quad}
\def\grafb#1{\epsfig{file=#1, width=7.7cm} \end{center}}
\def\be{\begin{equation}} \def\ee{\end{equation}}
\def\bea{\begin{eqnarray}} \def\eea{\end{eqnarray}}
\def\eq#1{(\ref{#1})}
\def\no{\nonumber}

\def\ep{\epsilon}
\def\pa{\partial}
\def\Z{\mathbb{Z}}

\def\cL{\mathcal{L}}

\def\cB{\mathcal{B}}
\def\cC{\mathcal{C}}
\begin{document}
\hfill SISSA 83/2002/FM 
\vskip 10mm
\sloppy

\title[Extended Toda Lattice]{Extended Toda Lattice}

\author{G. Carlet}
\address{Guido Carlet, SISSA, Via Beirut~2-4, 34014 Trieste, Italy}
\email{carlet@sissa.it}
\date{\today}

\begin{abstract}
We introduce nonlocal flows that commute with those of the classical Toda hierarchy. We define a logarithm of the difference Lax operator and use it to obtain a Lax representation of the new flows.
\end{abstract}
\maketitle

The present work is based on the joint work \cite{CDZ} of the author with B.~Dubrovin and Y.~Zhang.

The Toda lattice equation is a nonlinear evolutionary equation introduced by Toda \cite{toda} describing an infinite system of masses on a line that interact through an exponential force.
In suitable coordinates it can be written as a system
\bea
\frac{\pa}{\pa t} u_n &=& e^{v_{n+1}} - e^{v_n} \label{toda1}\\
\frac{\pa}{\pa t} v_n &=& u_n - u_{n-1}  \label{toda2} ,
\eea
with $n \in \Z$.

It was soon realized that this equation is completely integrable, i.e. admits infinite conserved quantities, can be solved for rapidly decreasing boundary conditions through the method of inverse scattering \cite{flaschka} and admits explicit quasi-periodic solutions by algebro-geometric methods \cite{DMN}-\cite{krich}. 
It has found important applications in many different fields, in particular in the theory of Gromov-Witten invariants of $C{\rm P}\sp 1$, where the present extension plays a particular role \cite{getzlercit}.

The Toda lattice equation can be seen as the first element of a whole hierarchy of commuting flows, the Toda lattice hierarchy. We can write all the flows by using the Lax representation \cite{flaschka}
\be
\ep \frac{\pa L}{\pa t_q} = [ \frac1{q+1!} (L^{q+1})_+ , L]
\ee
where $L$ is the following difference operator
\be
L= \Lambda + u +e^v \Lambda^{-1}  .
\ee
Here we are using a notation (see \cite{takasaki}) with a continuous space variable $x= n \ep$, where $\ep$ is the lattice spacing and $t_0 = \ep t$; i.e. $u(x) =u_n$ and $v(x) = v_n$ for $x =\ep n$. 
We denote by $\Lambda$ the shift operator $\Lambda f(x) = f(x+\ep)$. Given any difference operator $A= \sum_{k\in\Z} a_k \Lambda^k$ we denote by $A_+$, $A_-$ the positive and negative parts respectively, i.e. $A_+ = \sum_{k\geq 0} a_k \Lambda^k$, $A= A_+ + A_-$.
The first flow $t_0$ is simply given by the Toda equations \eq{toda1}-\eq{toda2}; as a further example the second flow is given by 
\bea
\ep u_{t_1}(x) &=& \frac12 (( u(x +\ep) + u(x)) e^{v(x+\ep)} - ( u(x-\ep) + u(x) ) e^{v(x)}) \\
\ep v_{t_1}(x) &=& \frac12 ( e^{v(x+\ep)} - e^{v(x-\ep)} + u^2(x)  -  u^2(x-\ep))  .
\eea

This hierarchy is Hamiltonian with respect to two different Hamiltonian structures \cite{kuper}; in our continuous formulation the first is given by the following Poisson bracket
\be
\label{pb1}
\{ u(x), v(y) \}_1 = \frac1{\ep} ( \delta(x-y+\ep) - \delta(x-y) )
\ee
and $\{ u(x), u(y) \}_1 = \{ v(x) ,v(y) \}_1 =0$.
The second Poisson bracket is 
\bea
&&\{ u(x), u(y) \}_2 = \frac1{\ep} (e^{v(x+\ep)} \delta(x-y+\ep) - e^{v(x)} \delta(x-y-\ep)) \\ 
&&\{ u(x), v(y) \}_2 =\frac1{\ep} u(x) (\delta(x-y+\ep ) - \delta(x-y)) \\
&&\{ v(x), v(y) \}_2 = \frac1{\ep} (  \delta(x-y+\ep ) - \delta(x-y-\ep )) . \label{pb4}
\eea
The equations of motion can be written in the following form
\be
\label{recrel}
\ep \frac{d}{dt_q} \cdot  = \{ \cdot , \bar{h}_q \}_1 = \frac1{q+1} \{ \cdot , \bar{h}_{q-1} \}_2 
\ee
where  $\bar{h}_q = \int h_q dx$ and the Hamiltonians are given as traces of powers of the Lax operator $L$ 
\be
\label{hami}
h_q = \frac1{(q+2)!} Res (L^{q+2}) ;
\ee
given any difference operator $A = \sum_{k\in\Z} a_k \Lambda^k$, its residue is defined by $Res A =a_0$.

The presence of a two Hamiltonian structures permits to obtain all the Hamiltonians through the Lenard-Magri \cite{magri} recursion relation \eq{recrel}
starting from a Casimir of the first Poisson bracket. 
In our case if we start from the Casimir $h_{-1} = u$ we obtain all the Hamiltonians defined above. Moreover this procedure guarantees that all the resulting Hamiltonians commute among themselves.

In order to see why and how this hierarchy of equations could be extended, let's consider its dispersionless limit that is obtained putting $\ep \to 0$. It can be shown \cite{takasaki} that the Lax representation of the dispersionless flows is given by 
\be
\frac{\pa \cL}{\pa t_{q}} = \{ \frac1{(q+1)!} ( \cL^{q+1} )_+  ,\cL \} ;
\ee
in this case $\cL$ is a function of $x$ and of the additional variable $p$
\be
\cL= p + u(x) + e^{v(x)} p^{-1}
\ee
and the  bracket is 
\be
\{ \cB, \cC \} = p \frac{\pa \cB}{\pa p} \frac{ \pa \cC}{\pa x} - p \frac{\pa \cC}{\pa p} \frac{\pa \cB}{\pa x} 
\ee
for $\cB$ and $\cC$ functions of $p$ and $x$.
$(\cB)_+$ means that only non-negative powers of $p$, in the power series expansion of $\cB$, are considered.

The dispersionless Hamiltonians and the Poisson brackets are simply obtained from their dispersive counterparts \eq{hami}, \eq{pb1}-\eq{pb4} by putting $\ep \to 0$.
In particular the same recursion relation as above \eq{recrel} holds in the dispersionless case.

Considering the genus zero approximation of the topological $C{\rm P}\sp 1$ model, in \cite{eguchi} it was noted that new flows can be added to the usual dispersionless flows given above; their Lax representation is
\be
\label{laxdispless}
\frac{\pa \cL}{\pa \tilde{t}_{q}} = \{ \frac2{q!} ( {\cL}^q (\log {\cL} - c_q) )_+ , \cL \} 
\ee
where $c_q = \sum_{k=1}^q \frac1k$, $c_0=0$.
The logarithm of $\cL$ must be understood in the following way 
\be
\log {\cL} = \frac12 v + \frac12 \log(1+ u p^{-1} +e^v p^{-2} ) + \frac12 \log ( 1+ u e^{-v} p +e^{-v} p^2)
\ee
where the first logarithm  on the RHS is seen as an expansion in negative powers of $p$ while the second one in positive powers of $p$.

These flows can be expressed in Hamiltonian form by
\be
\frac{d}{d \tilde{t}_q} \cdot = \{ \cdot , \bar{\tilde{h}}_q^{disp} \}_1
\ee
where the Poisson bracket is the dispersionless limit of \eq{pb1} and the dispersionless Hamiltonians are given by
\be
\tilde{h}_q^{disp} = \frac2{(q+1)!} Res_{p=0} [ p^{-1}  \cL^{q+1} ( \log \cL - c_{q+1}) ] .
\ee
These Hamiltonians however satisfy a recursion relation that is different from the previous one \eq{recrel}
\be
\label{recrel1}
\{ \cdot , \bar{\tilde{h}}_{q-1} \}_2 = q \{ \cdot ,  \bar{\tilde{h}}_q \}_1 + 2 \{ \cdot , \bar{h}_{q-1} \}_1 .
\ee

We would like to briefly mention that in the dispersionless limit all the flows can be introduced in a rather different fashion, by using the relation between systems of hydrodynamic type and Frobenius manifolds \cite{dubrovin-flat}.
The dispersionless Toda Poisson pencil is associated to a Frobenius manifold characterized by the free energy $F=\frac12 u^2 v + e^v$; then all the Hamiltonians  of the system can be obtained by an expansion of the so-called deformed flat coordinates of the Frobenius manifold. The relationship between the dispersive and dispersionless versions of the Toda hierarchy is just a particular instance of the classification program of bihamiltonian integrable hierarchies proposed in \cite{DZ}, based on the reconstruction of the whole dispersive hierarchy starting from its dispersionless limit.

Thus we have seen that in the dispersionless case the Toda hierarchy has two perfectly well defined sequences of flows all commuting between themselves, denoted by the times $t_q$ and $\tilde{t}_q$ for $q \geq 0$.
The classical dispersive flows corresponding to the times $t_q$ defined above reduce, for $\ep \to 0$, to the corresponding flows in the dispersionless hierarchy; on the other hand  in the classical dispersive formulation there is apparently no flow reducing for $\ep \to 0$ to the dispersionless flows corresponding to the times $\tilde{t}_q$.

In analogy with the Lenard-Magri procedure for the first set of Hamiltonians one expects  to find the second set of flows of the dispersive hierarchy from another Casimir of the first Poisson bracket by applying the recursion relation \eq{recrel1} (this time with the full dispersive brackets). The first Poisson bracket admits in fact a second Casimir: $\tilde{h}_{-1} = v$. However the recursion relation \eq{recrel1} fails to work in this case, as one can easily check, the reason being that  $\tilde{h}_{-1}$ is a Casimir of {\it both} Poisson brackets.
This phenomenon is referred to as ``resonance'' of the Poisson pencil.

We can however introduce an ansatz for the first nontrivial Hamiltonian $\tilde{h}_0$; 
it was given in \cite{zhang} and \cite{getzlerans} and is such that the corresponding flow coincides with the $x$-translation
\be \label{ansatz}
\tilde{h}_0 =  u \Lambda (\Lambda -1)^{-1} \ep v_x .
\ee
This Hamiltonian is nonlocal, since it contains the inverse of the discrete derivative; this can be written as a formal series in $\ep$ 
\be
(\Lambda -1)^{-1} \ep v_x = \sum_{k \geq 0} \frac{B_k}{k!} (\ep \pa_x)^k v
\ee
where the Bernoulli numbers $B_k$ are defined by $\frac{x}{e^x -1} = \sum_{k \geq 0} \frac{B_k}{k!} x^k$. 
Starting from this ansatz one can define all the Hamiltonians $\tilde{h}_q$ using the recursion relation \eq{recrel1} and show that they commute between themselves and with all classical Toda flows.

To have an explicit form for these new flows, it is important to find their Lax representation.
By considering the dispersionless Lax representation \eq{laxdispless} one expects that it will be necessary to introduce a logarithm of the Lax operator $L$; such operator can be defined by the dressing formalism.
It is well known \cite{ueno} that one can write $L$ as the dressing of the shift operators $\Lambda$ and $\Lambda^{-1}$
\be \label{dress}
L = P \Lambda P^{-1} = Q \Lambda^{-1} Q^{-1} 
\ee
where 
\bea
&&P = \sum_{k \geq 0}  p_k \Lambda^{-k} \qquad p_0 =1 \\
&&Q = \sum_{k \geq 0} q_k \Lambda^k .
\eea
By substituting in the definition \eq{dress}, the functions $p_k$, $q_k$ can be found in terms of $u$, $v$.

Noticing that $\Lambda = e^{\ep \pa_x}$ one is led to define two different logarithms in the following way
\bea
&&\log_+ L = P \ep \pa P^{-1} = \ep \pa + P \ep P^{-1}_x  \\
&&\log_- L = -  Q \ep \pa Q^{-1} = - \ep \pa -  Q \ep Q^{-1}_x .
\eea
These logarithms are both differential--difference operators, due to the presence of $\ep \pa$;
since we want to write an expression like \eq{laxdispless} and we need to make sense of the $(\cdot)_+$  part, we would like to have a purely difference operator for the logarithm, that we define by 
\be
\log L = \frac12 \log_+ L + \frac12 \log_- L = -\frac{\ep}2 ( P_x P^{-1} -  Q_x  Q^{-1}) ;
\ee
in this definition the derivative drops out and we get a difference operator of the form
\be
\log L = \sum_{k \in \Z} w_k \Lambda^k .
\ee

One would like to express the coefficients $w_k$ in terms of the variables $u$, $v$. This is indeed possible (see \cite{CDZ} for a proof);
essentially, by definition, all the previously defined logarithms commute with $L$, e.g.
\be \label{commplus}
[ \log_+ L , L ] =0 ;
\ee
inserting
\be
\log_+ L = \ep \pa + 2 \sum_{k \leq -1} w_k \Lambda^k
\ee
into \eq{commplus} one can solve it recursively and express the coefficients $w_k$ for $k\leq -1$ as formal power series in $\ep$ with coefficients that are differential polynomials in $u$ and $v$. The analogous expression for $\log_- L $ gives the coefficients $w_k$ for $k \geq 0$. First few examples are
\bea
w_{-1} &=&\frac12 (\Lambda-1)^{-1} \ep u_x \\
w_0 &=&\frac12 \Lambda ( \Lambda -1)^{-1} \ep v_x \\
w_1 &=&\frac12 \Lambda e^{-v} (\Lambda -1)^{-1} \ep u_x .
\eea
Observe that this is in general not possible for the coefficients $p_k$, $q_k$ of the dressing operators.

Using this definition of logarithm of $L$ we can give the Lax representation for the new flows, which is formally analogous to the dispersionless Lax representation
\be
\ep \frac{\pa L}{\pa \tilde{t}_q} = [ A_q , L ] \qquad A_q = \frac2{q!}[ L^q ( \log L -c_q)]_+  ; 
\ee
here $A_q$ is a difference operator of infinite order; equivalently we can use the operator
\be
\tilde{A}_q =  \frac2{q!}[ L^q ( \log L -c_q)]_+  - \frac1{q!}  L^q (\log_- L -c_q)
\ee
that also contains a differential part but is of finite order; it gives the same Lax equations since it differs from $A_q$ by a part that commutes with $L$. For example, the first two Lax operators are given by 
\bea
\tilde{A}_0 &=& \ep \pa   \\
\tilde{A}_1 &=& \Lambda (\ep \pa -1) + \Lambda (\Lambda -1)^{-1} \ep u_x + u ( \ep \pa -1) \\ 
&&+ e^{v} ( \ep \pa +1 - (\Lambda -1)^{-1} \ep v_x) \Lambda^{-1} . \no
\eea
As expected the first one corresponds to the $x$-translations while the second gives the first nontrivial extended Toda flow
\bea
\label{flow-t11-1}
\ep u_{\tilde{t}_{1}} &=& (\Lambda-1)(- e^{v} (\Lambda^{-1}-1)^{-1} \ep  v_x ) -2 (\Lambda -1 ) e^{v} \\
&& + \ep (\frac{u^2}2  + e^v )_x   \no \\ 
\label{flow-t11-2}
\ep v_{\tilde{t}_{1}} &=& ((\Lambda^{-1} -1)^{-1} \ep v_x )(\Lambda^{-1}-1)u +\ep v_x (\Lambda^{-1} u) \\
&&+\Lambda^{-1} \ep u_x +\ep u_x + 2 (\Lambda^{-1} -1) u . \no
\eea

We can also give an explicit expression for the Hamiltonians in analogy with the dispersionless case
\be
\tilde{h}_q = \frac2{(q+1)!} Res [ L^{q+1} (\log L - c_{q+1}) ] ;
\ee 
these are exactly the Hamiltonians defined above by the recursion relation and the ansatz \eq{ansatz}, up to total derivatives.

The new flows of this extended Toda hierarchy are nonlocal but nevertheless have many of the nice properties that are expected from a completely integrable system, such as the existence of multisolitonic solutions; for example, the evolution of a simple soliton under the nonlocal flows \eq{flow-t11-1}-\eq{flow-t11-2} is given by 
\be
u(x,\tilde{t}_1) = (\Lambda -1) \frac{\cosh (\frac1{\ep} ((x+ \lambda_1 \tilde{t}_1) \log z_1 + \tilde{t}_1 (-z_1 +1/z_1)))}{\cosh (\frac1{\ep} ((x+ \lambda_1 \tilde{t}_1) \log z_1 + \tilde{t}_1 (-z_1 +1/z_1))- \log z_1 )} \no
\ee
\be
v(x, \tilde{t}_1) = (1-\Lambda^{-1})^2 \log [ 2 \cosh( \frac1{\ep} (x + \lambda_1 \tilde{t}_1) \log z_1 + \frac1{\ep} \tilde{t}_1 (-z_1 +1/z_1))  ]  \no
\ee
where $\lambda_1 = z_1 + z_1^{-1}$ and $z_1$ is a parameter. As usual this is simply obtained by Darboux transformations of a particular constant solution.

One expects that also the algebro-geometric quasi-periodic solutions should have a nice behavior under these flows; this problem is however still under consideration. Further generalization for the multicomponent case is in progress.

All the proofs of the above statements will appear in \cite{CDZ}.

\end{document}